# Towards Brain-inspired Computing

Zoltan Gingl [x,y], Sunil Khatri [+] and Laszlo B. Kish [+]

[x] *Department of Experimental Physics, University of Szeged, Dom ter 9, Szeged, H-6720 Hungary*

[+] *Department of Electrical and Computer Engineering, Texas A&M University, College Station, TX 77843-3128, USA*

(1st version, March 20, 2010)

**Abstract.** We present introductory considerations and analysis toward computing applications based on the recently introduced deterministic logic scheme with random spike (pulse) trains [*Phys. Lett. A* **373** (2009) 2338-2342]. Also, in considering the questions, "Why random?" and "Why pulses?", we show that the random pulse based scheme provides the advantages of realizing multivalued deterministic logic. Pulse trains are realized by an element called orthogonator. We discuss two different types of orthogonators, parallel (intersection-based) and serial (demultiplexer-based) orthogonators. The last one can be slower but it makes sequential logic design straightforward. We propose generating a multidimensional logic hyperspace [*Physics Letters A* **373** (2009) 1928–1934] by using the zero-crossing events of uncorrelated Gaussian electrical noises available in the chips. The spike trains in the hyperspace are non-overlapping, and are referred to as neuro-bits. To demonstrate this idea, we generate 3-dimensional hyperspace bases using 2 Gaussian noises as sources for neuro-bits, respectively. In such a scenario, the detection of different hyperspace basis elements may have vastly differing delays. We show that it is possible to provide an identical speed for all the hyperspace bases elements using correlated noise sources, and demonstrate this for the 2 neuro-bits situations. The key impact of this paper is to demonstrate that a logic design approach using such neuro-bits can yield a fast, low power processing and environmental variation tolerant means of designing computer circuitry. It also enables the realization of multi-valued logic, significantly increasing the complexity of computer circuits by allowing several neuro-bits to be transmitted on a single wire.

## 1. Introduction.

Recently, a new deterministic, multivalued logic scheme based on the functioning of the brain was introduced [1]. The logic utilized random spike (pulse) trains and a hyper-space scheme [2] similar to the quantum Hilbert space was developed. The random spike trains corresponding to different hyper-space elements are non-overlapping (orthogonal). This was shown to outperform a quantum search algorithm [2]. The hyperspace in [2] was a utilization of the generic hyperspace elements introduced for noise-based logic [3] with continuum noises, and it was shown to be as rich as a carrier of information as the quantum Hilbert space [2]. These results are partial answers to questions about whether computing hardware could be driven by noise in order to reduce energy dissipation [4].

In this paper we make the first step toward the realization of digital logic using the scheme introduced in [1]. First we discuss why spike-based logic is feasible, and how random spike train based logic signals can be beneficial.

---

[y]   Corresponding author.

Bollapalli et al [5] used two non-orthogonal sinusoidal signals to represent logic states, and demonstrated the realization of AC-voltage-based (binary) logic. This is the first step toward a multi-valued logic system, where the base vectors are orthogonal sinusoidal signals with different frequencies and/or $\pi/2$ phase shift. The advantage of this sinusoidal scheme is that in the binary valued logic case it is about 100 times faster than the noise-based binary logic scheme [3], provided that the amplitude of the sinusoids are well beyond the background noise. However, the last condition guarantees that the sinusoidal scheme [5] cannot offer the lowest possible power dissipation.

In this paper, we demonstrate that random spike trains can be practically utilized to implement logic. We demonstrate how orthogonal spike trains can be derived from random (possibly overlapping) source spike trains, using a simple circuit called an orthogonator. We demonstrate two kinds of orthogonators (demultiplexer-based and intersection-based) and show that by using intersection-based orthogonators and N random spike trains, we can generate an exponentially larger hyperspace basis of orthogonal spike trains. Each such spike train represents a neuro-bit. The first coincident spike of any neuro-bit can identify it, resulting in a fast and resilient logic family. In recent times, computer circuits are plagued with the problem of processing variations [6]. Using the concepts outlined in this paper, variation tolerant circuits can be designed, while speed is retained. Furthermore, the logic approach described in this paper makes it easy to implement multi-valued logic functions, something that traditional digital VLSI design simply cannot achieve in practice. Additionally, the logic approach of this paper lends itself to extremely low power design using sub-threshold operation [7, 8], on account of its high resilience to processing and environmental variations (which are aggravated in sub-threshold design).

The remainder of this paper is organized as follows. Section 2 describes why random spikes are utilized in our logic approach, illustrating the resilience and speed that can result from this choice. The key circuit required to realize this new logic approach is the orthogonator. Section 3 describes the design of two kinds of orthogonators. Section 4 presents results from initial experiments that we have conducted to compare the outputs of the two orthogonators. We discuss the properties of the resulting neuro-bits, and show how such orthogonators may be used to generate hyperspace elements. In Section 5, we outline how elementary gates and set operations are performed using random spike trains. Section 6 elaborates on the reason why a stochastic spike train is superior to a periodic one, while Section 7 concludes the paper.

**2. Why random and why spikes?**

At the moment, there is no complete answer to the questions if periodic signals or noise should be used to generate the pulse sequences and if spikes are the most advantageous. We have some partial answers below. The fact that the brain uses stochastic spike trains may be a further indication that this is the way to go.

However, there is an answer for the case of the ultimate lower limit of energy dissipation to process a bit, assuming ideal devices for amplification and switching – or in other words, controlled potential barriers without parasitic elements. In this case, the lowest energy dissipation for a single gate is reached by noise because the "noise clock" signal can be dissipation-free, using simply the thermal noise of a resistor distributed in a frequency-dispersion-free line [4]. In order to keep power dissipation at a minimum, the amplifier stages for making the local reference basis signals will amplify this noise so that a given amplification stage has just barely enough supply voltage to handle that amplitude of noise. The subsequent stages with greater noise signal amplitudes will use correspondingly greater supply voltages. The zero crossings events of the amplified noise make the local hyperspace reference vectors. On the other hand, using a periodic clock signal means increased energy dissipation.

Furthermore, the spike-noise based logic scheme [1] has an important advantage compared to continuum-noise-based logic with superposition of orthogonal elements: to determine a logic value, that is, to correlate it with the different reference base values, the spike-based scheme does not need time averaging and therefore results in a significant speed-up. Due to the orthogonality of base (reference) spike trains, simple coincidence detection of a single spike can identify any reference spike train uniquely[1]. This property, when utilized with a multivalued logic scheme, may be the key explanation for the fact that the brain can perform efficiently with slow and random spike trains and can recognize/analyze complex situations very fast.

Finally, in answer to the question: why stochastically timed spike trains are present (like in the brain) instead of periodically timed spike trains, we show that random spike trains are more resilient to circuit delays, while periodic spike trains are particularly susceptible to delays that may arise from processing and environmental variations. This discussion is continued in further detail in Section 6. We conclude in Section 7, with some suggestions for future work in this area.

**3. Orthogonator types**

To construct the *M*-dimensional orthogonal bases of the hyperspace, we need orthogonal spike trains. In [1], the base elements were generated from partially overlapping random neural spikes by a neural circuit called an *orthogonator*. The order of the orthogonator can be related to *M*, since an *N*-th order orthogonator generates $M = 2^N - 1$ outputs. Here we will call the orthogonator circuit of [1] as a "*intersection-based orthogonator*" and, in the present paper, we present a new kind of orthogonator as follows:

i) *Demultiplexer-based orthogonator*: It has a single input which is fed by a single (infinite) spike sequence. In an M-th order orthogonator a *demultiplexer* will distribute the subsequent spikes to *M* output wires in a cyclic way as follows:

$$p = 1 + (r-1) \mod M \quad ,$$

where $p$ ($1 \leq p \leq M$) is the index number of the output wire on which the *r*-th input pulse will emerge. The resulting spike trains in the *M* separate output wires are orthogonal by construction. Moreover, when the *M*-th wire outputted its *k*-th spike, we know that the previous *M*-1 spikes were outputted on the other *M*-1 wires (one spike for each wire). All the M spikes mentioned above belong to the *k*-th spike package of size *M* of the original spike train. Thus the advantage of a demultiplexer-based orthogonator is that it makes easy/natural to construct sequential logic operations and networks. Each such spike train (of any wire) represents one element of an *M*-dimensional reference basis $\{V_i(t_k)\}$ (i=1...*M*) at a given "computer time" $t_k$ (even though the physical time moments of the actual spikes in the *k*-th spike package are different). Another advantage is that the average pulse rate on the output wires is identical.

ii) *Intersection-based orthogonator*: Such a circuitry [1] has $N$ parallel inputs driven by parallel, partially overlapping random spikes trains. The orthogonator generates all combinations of the available set-theoretical intersections of the input spikes. As a result there are $M = 2^N - 1$ output wires with non-overlapping spike trains [3]. The advantage of the intersection-based orthogonator is that it may be faster and it can transform a set of partially overlapping spike trains into a set of orthogonal spike trains. On the other hand, the construction of sequential logic operations and networks from spike trains generated from intersection-based orthogonators is less obvious and to provide similar pulse rates on the output wires is a nontrivial task, see below in Section 4.2.

**4. Demonstration: generating the neuro-bits and hyperspace base using logic circuits**

In this section, we compare the results obtained by demultiplexer-based and intersection-based orthogonators, using computer simulations. We compare 1/f and white noise as possible sources of the noise spike trains, and also show how the properties of the hyperspace basis elements vary (in terms the mean and rms of fluctuations of the inter-spike intervals in all cases).

*4.1 Generation by a demultiplexer-based orthogonator*

The demonstration of a second-order demultiplexer-based orthogonator can be seen in Figure 1. The random spikes generated by the zero crossing events of a band-limited white noise (top plot of Figure 1) are cyclically demultiplexed to three output wires (bottom 3 plots of Figure 1).

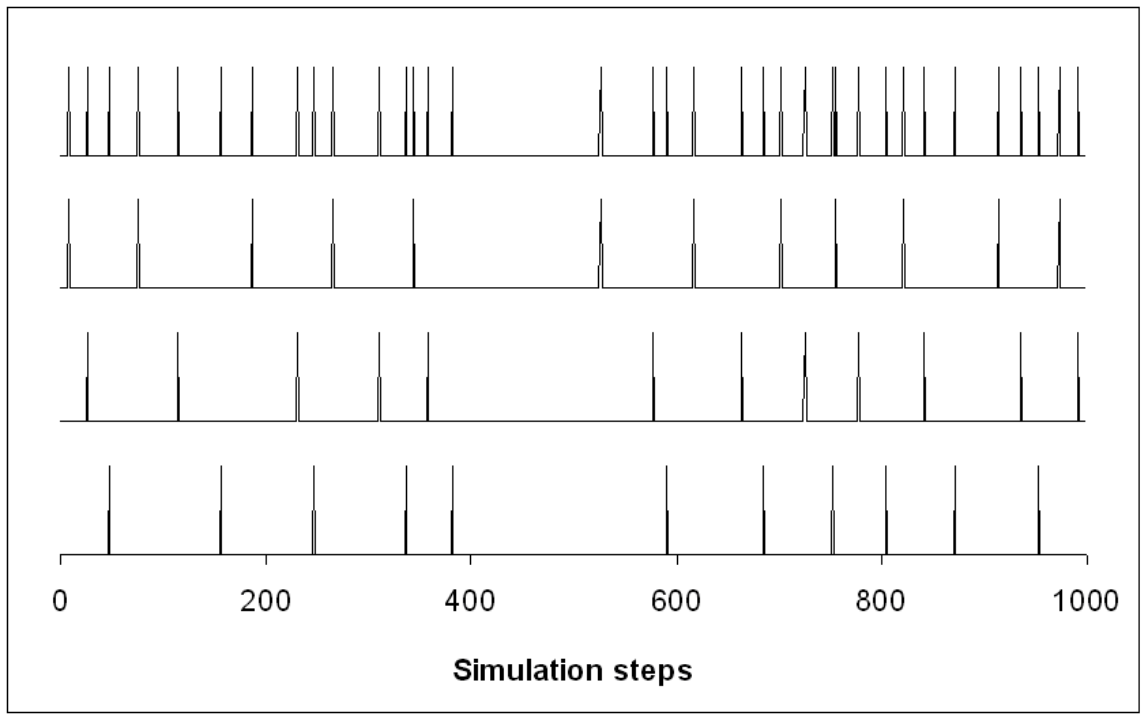

**Figure 1.** Second-order demultiplexer-based orthogonator driven by zero-crossing events of band-limited white noise. Upper plot: the original spike train. Lower plots: the orthogonal sub-trains at the three outputs.

In Table 1, the statistical values in a second-order demultiplexer-based orthogonator, the mean interspike intervals ($\tau$) and their *rms* fluctuation values ($\Delta\tau$) are shown for the case of band-limited white noise and band-limited 1/f noise. It is obvious from the data in Table 1 that spike trains generated by white noise have superior properties compared to 1/f noise in regarding the mean frequency of spikes and the fluctuations of their locations whenever these quantities matter.

| S(f) | $\tau$, source spike train | $\Delta\tau$, source spike train | $\tau$, output spike trains | $\Delta\tau$, output spike trains |
|---|---|---|---|---|
| White (5MHz -10GHz) | 90ps | 58ps | 267ps | 100ps |
| 1/f (2.5MHz -10GHz) | 225ps | 469ps | 681ps | 768ps |

**Table 1.** Results of statistics with the second-order demultiplexer-based orthogonator, using 65536 simulation points. The numeric frequency and time data scaled up to practical values.

### *4.2 Base generation by a intersection-based orthogonator and homogenizing their outputs*

In Figure 2, a second-order intersection-based orthogonator is demonstrated. Two band-limited Gaussian white base noises used to generate the original spike trains by their zero-crossing events. The upper two plots show the original input spike trains of the two inputs A and B. The lower three plots correspond to the three orthogonal outputs: $AB$, $A\overline{B}$ and $\overline{A}B$. It is obvious that the pulse frequencies of different output types ($AB$ versus $A\overline{B}$ and $\overline{A}B$) are very different, which can be disadvantageous for practical applications because different bit values (basis spike trains) may have different speed since the first pulse in these bases occurs at widely varying time instants. However, we note that some applications may prefer to use the fastest possible rate for high-value bits and lower speed at low-value bits as described below. This feature can be utilized to advantage in practical applications. Also, note that in all cases, using *N* original spike trains can result in $2^N - 1$ basis elements.

The difference in pulse frequencies between the derived pulse trains can however be reduced or eliminated by using correlated noises to generate the input spike trains. Such an operation is able to "homogenize" spike frequencies at the different outputs while the orthogonality is maintained. Figure 2 shows example simulation data for homogenized spike frequencies of a second-order orhogonator. The correlation between the two noises generating the input spikes is generated by using a third common mode noise, which is added to the original noises. The noise amplitude of the common noise is 0.945, while the non-correlated noises have an amplitude of 0.055.

It is interesting to note that without homogenization, the slow (*AB*) bit can be used for the lower bit values and the faster ones for the higher values. Thus, in a short time, coincidences between the signal spikes and the fast reference trains' spikes will quickly provide a rough output. Then the accuracy of the output will gradually be refined by subsequent coincidences between the signal spikes and spikes of the relevant low-bit-value reference spikes.

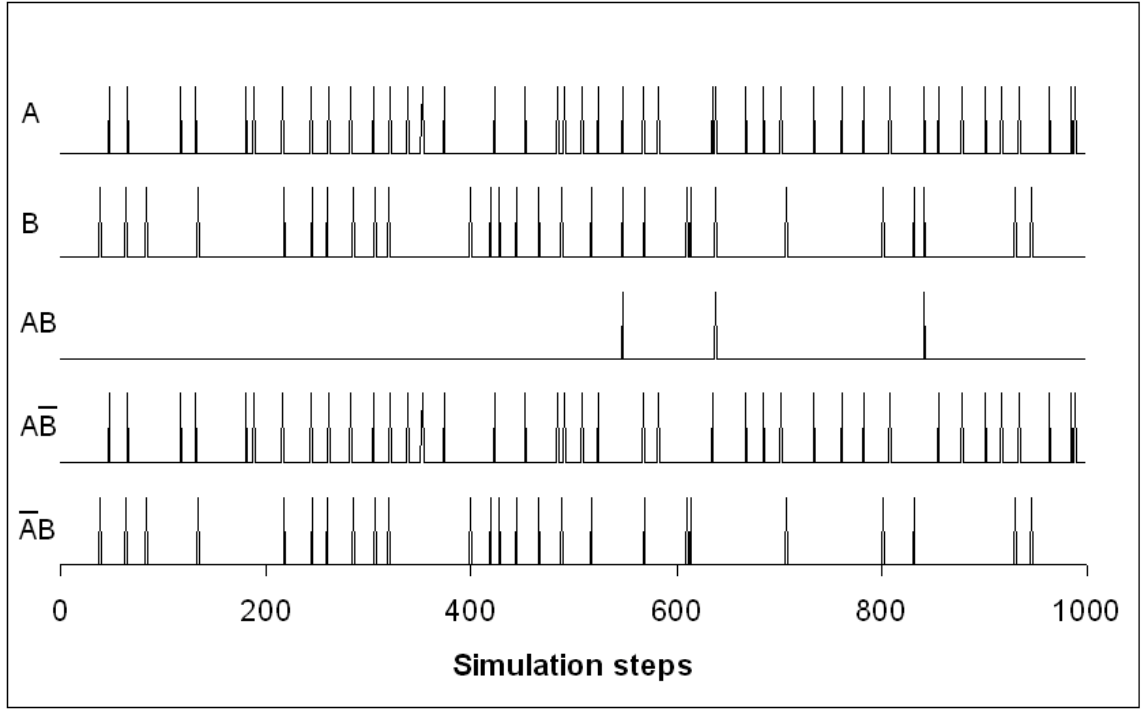

**Figure 2**. Second-order intersection-based orthogonator driven by the zero-crossing events of two independent band-limited white noises. Upper two plots: the original input spike trains, lower plots: the three different types of orthogonal sub-trains at three outputs.

|  | Noncorrelated source | | Correlated source | |
|---|---|---|---|---|
|  | $\tau$ | $\Delta\tau$ | $\tau$ | $\Delta\tau$ |
| $A$ | 28<br>90ps | 18<br>58ps | 28<br>90ps | 19<br>61 |
| $B$ | 28<br>90ps | 19<br>61ps | 28<br>90ps | 19<br>61ps |
| $AB$ | 697<br>2.24ns | 678<br>2.18ns | 52<br>167ps | 46<br>148ps |
| $A\overline{B}$ | 29<br>93ps | 20<br>64ps | 58<br>186ps | 53<br>170ps |
| $\overline{A}B$ | 30<br>96.4ps | 21<br>67.5ps | 59<br>190ps | 54<br>174ps |

**Table 2.** Computer simulation results on the second order intersection-based orthogonator driven by non-correlated or specifically correlated noises in order to homogenize pulse rates, using 65536 points. Simulated numeric frequency data scaled up to practical values.

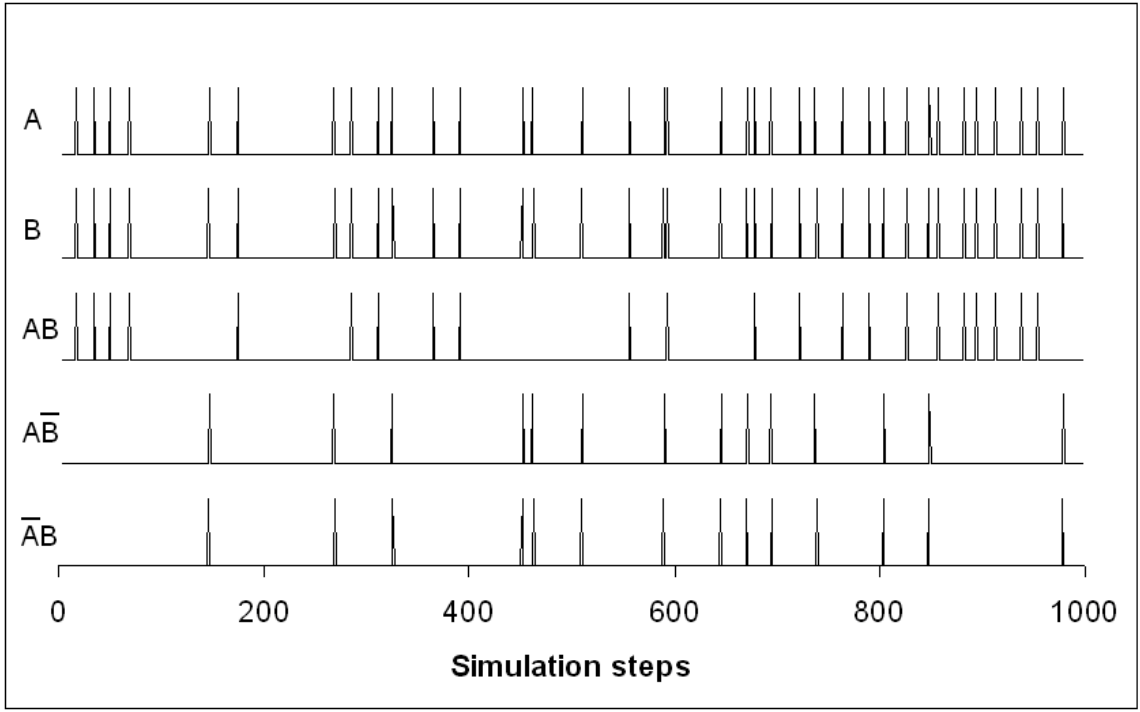

**Figure 3**. Second-order intersection-based orthogonator driven by the zero-crossing events of two strongly correlated band-limited white noises. Upper two plots: the original input spike trains, lower plots: the three different types of orthogonal sub-trains at three outputs.

In Table 2, the statistical values in the intersection-based second-order orthogonators (Figures 2 and 3), the mean interspike intervals ($\tau$) and their *rms* fluctuation values ($\Delta\tau$) are shown for the case of non-correlated (Figure 2) and correlated (Figure 3) input noises.

**5. Common Logic Operations using Random Spike Trains**

Computation using the random spike trains proceeds along the lines of [1], [2]. Assuming we have *M* orthogonal spike trains, each of these *M* basis elements can be treated as a point in a multi-variable space. Assume that there are *K* input wires (1, 2, … i … K) for a gate, each with $M_i$ possible values, where $\prod M_i = M$. Then, elementary set operations are performed using the ideas of [1]. In this case, we assume that all values of the *K* input wires are members of the same hyperspace. In the case of elementary logic gates, the gate operations proceed along the lines of [2]. In this case, the gates have correlators for each input, which determine the value of the input in a multi-variable space. Based on the input values, the gate drives out an appropriate output, possibly from a different hyperspace than the hyperspace of the inputs.

In both the above discussions, the operations are performed extremely quickly since the first coincident spike of any neuro-bit can identify it, resulting in an extremely fast logic family. Thus elementary gate operations (complementation, logical AND, logical OR etc) or elementary set operations (membership tests, set union or intersection) can be done extremely fast even though the hyperspace is extremely large. We conjecture that the brain may be using such a logic approach, allowing it to do many complex reasoning and recognition operations extremely fast.

## 6. Why noise spikes and why not periodic?

Here we present an answer to the question: why should we use stochastically timed spike trains generated by a demultiplexer-based orthgonator (like in the brain) instead of periodically timed spike trains? Such a periodic arrangement would provide the best filling with the uniform spike trains having the highest spike frequency.

The answer is straightforward: In the periodic case, the generated orthogonal (that is, non-overlapping) periodic spike sequences would have the same pattern; they would be the time-shifted versions of each other. Thus, two different basis elements would result in aliasing if one of them is appropriately delayed, resulting in an unreliable circuit. Furthermore, generating periodic spike trains in reality is harder, since random spike trains are more natural. While we are using the noisy spike train, each orthogonal vector represents a unique fingerprint, with a minimal likelihood of aliasing among different basis elements. This property appears to be a fundamentally important feature of noise-based logic.

## 7. Conclusions

In this paper, we have described a logic approach based on a hyperspace constructed from orthogonal spike trains. The key strength of this approach is its significant speed, high resilience and its ability to implement multi-valued logic. We believe that these features allows this approach to be a key enabling approach for a new class of computing circuits, which can address the problems being faced by the traditional digital design approaches in use today. In the future, we plan to design digital circuits using this approach, and compare the resulting circuits with existing design methodologies.

## Acknowledgements

This work was supported by grant OTKA K69018 of the Hungarian Academy of Sciences. The authors thank Robert Mingesz for useful discussions.